\documentclass[twocolumn,DIV=17,fontsize=10pt,headinclude,abstract]{scrartcl}

\pdfoutput=1

\recalctypearea

\usepackage[utf8]{inputenc}
\usepackage[T1]{fontenc}
\usepackage[english]{babel}

\usepackage{scrlayer-scrpage}
\lohead{RIOT-POLICE}
\rohead{Sören Tempel and Tristan Bruns}

\usepackage[tt=false]{libertine}
\usepackage[binary-units]{siunitx}
\usepackage{hyperref}
\hypersetup{
	colorlinks=true,       
	linkcolor=blue,        
	citecolor=blue,        
	urlcolor=blue          
}

\usepackage{csquotes}

\usepackage[
	abbreviate=true,
	dateabbrev=true,
	isbn=true,
	doi=true,
	urldate=comp,
	url=true,
	maxbibnames=2,
	maxcitenames=2,
	backref=false,
	backend=biber,
	style=numeric,
	language=american,
]{biblatex}

\begin{document}
	\title{RIOT-POLICE: An implementation of spatial
		memory safety for the RIOT operating system}

	\author{%
		Sören Tempel\\%
		University of Bremen\\%
		tempel@uni-bremen.de%
		\and%
		Tristan Bruns\\%
		University of Bremen\\%
		tbruns@uni-bremen.de}

	\date{September of 2018}

	\hypersetup{
		pdftitle={RIOT-POLICE},
		pdfauthor={Sören Tempel and Tristan Bruns}
	}

	\maketitle

	\begin{abstract}
		We present an integration of a safe C dialect, Checked
		C, for the Internet of Things operating system
		RIOT. We utilize this integration to convert parts of
		the RIOT network stack to Checked C, thereby
		achieving spatial memory safety in these code parts.
		Similar to prior research done on IoT operating systems
		and safe C dialects, our integration of Checked C
		remains entirely optional, i.e. compilation with a
		standard C compiler not supporting the Checked C
		language extension is still possible. We believe this to
		be the first proposed integration of a safe C dialect
		for the RIOT operating system. We present an incremental
		process for converting RIOT modules to Checked C,
		evaluate the overhead introduced by the conversions,
		and discuss our general experience with utilizing
		Checked C in the Internet of Things domain.
	\end{abstract}

	\section{Introduction}

RIOT is an operating system explicitly targeting \enquote{constrained
IoT devices} \cite[2453]{baccelli2013riot}. As such these devices are
programmed in the low-level programming language C
\cite[2454]{baccelli2013riot} which offers very few safety features.

The result is that errors in programs written in C often go unnoticed
\cite[103]{devietti2008hardbound}. This is especially critical in the
Internet of Things for a variety of reasons. First of all, debugging of
constrained IoT devices is difficult \cite[1]{levy2015tock}. If a bug
has been found, debugged and fixed an update needs to be provided.
Unfortunately, updating constrained devices is challenging
\cite{ietf-suit-information-model-01} and at the time of writing RIOT
doesn't provide any built-in method for doing so
\cite{acosta2017updates}. Additionally, some programming errors caused
by missing safety features can be exploited. Buffer overflows are the
standard example for this \cite{cowan2000bufferoverflows}. This is even
more problematic on constrained IoT devices where protection mechanisms
such as fault isolation are not available
\cite[234]{levy2017multiprogramming}.

For this reason, it is important to prevent such errors in the programming
language itself. Doing so has been a longstanding research topic
\cite{devietti2008hardbound, grossman2005cyclone, zhou2006safedrive,
necula2002ccured}. One of the more recent research projects in this
regard is an extension of the C programming language called Checked C
\cite{tarditi2017checkedc}. In this paper we will present an
integration of the Checked C language extension for the RIOT operating
system. With this integration we are hoping to bring spatial memory
safety to the RIOT operating system.
To our knowledge, this is the first work exploring Checked C in the
embedded domain. Our integration of Checked C remains entirely optional,
thereby allowing utilization of Checked C without loosing the ability
to compile RIOT with a legacy C compiler. 

\section{Background}

The following subsections briefly introduce the technologies used in
this paper. Related work is also presented.

\subsection{Checked C}

Checked C is an extension of the C programming language. It extends the
C programming language as standardized by the ISO \cite{ISO9899}
(referred to as legacy C in the following text) with facilities to make
writing spatial memory safe C code possible. Spatial memory safety
ensures that \enquote{any pointer dereference is always within the
memory allocated to that pointer} \cite[1]{tarditi2017checkedc}.

Since it introduces new syntactic keywords, Checked C code needs to be compiled
with a custom compiler based on LLVM/Clang. It cannot be compiled with a
standard legacy C compiler such as GCC. Nonetheless, it has a strong
focus on backwards compatibility allowing developers to call legacy C
code from Checked C code and vice versa \cite[2]{tarditi2017checkedc}.
Additionally, evaluations of the code emitted by the Checked C compiler
have shown that the language extension introduces a comparably low
executable size of \SI{7.4}{\percent} on average
\cite[10]{tarditi2017checkedc}. This makes Checked C especially well
suited for constrained IoT devices where program memory is
often limited \cite[2453]{baccelli2013riot}.

\subsection{RIOT OS}

RIOT OS is an open source real-time operating system for the Internet of
Things. It supports constrained IoT devices with at least
\SI{1.5}{\kilo\byte} of RAM and \SI{5}{\kilo\byte} of ROM. Nonetheless,
it has features known from conventional operating systems. For
instance, it supports multi-threading and provides a network stack
\cite[2454]{baccelli2013riot}.

Compared to other operating systems for the Internet of Things, for
example Contiki \cite{dunkels2004contiki} or TinyOS \cite{levis2005},
RIOT is written entirely in standard ISO C \cite[2454]{baccelli2013riot}.
As such, it is not memory safe and subject to bugs caused by memory
safety violations such as buffer overflows. Unsurprisingly, out-of-bounds buffer
accesses have been found in RIOT and fixed in the past
\cite{lenders2016pktbuf, lenders2017ipv6addr, willemsen2017cbor,
tempel2018ieee802154}.
Unfortunately, the RIOT developers don't gather these issues in a
central place. Therefore, it is difficult to generate statistics for
these kinds of bugs.

\subsection{Related work}

Various attempts have been made in the past to bring memory safety to
different operating systems for the Internet of Things.
\citeauthor{cooprider2007tinyos} tried to achieve memory safety by
integrating Deputy into the TinyOS ecosystem \cite{cooprider2007tinyos}
(referred to as Safe TinyOS in the following text). A similar approach,
also utilizing Deputy, was implemented by
\citeauthor{dunkels2004contiki} for Contiki
\cite{dunkels2004contiki} (referred to as Safe Contiki OS in the
following text). Especially the former research lead to some
interesting results indicating a ROM overhead of \SI{13}{\percent} and a
CPU overhead of \SI{5.2}{\percent} on average
\cite[2]{cooprider2007tinyos}.

To our knowledge no attempts haven been made to bring similar techniques
to the RIOT operating system. Whether or not this is possible is subject
of this paper. Contrary to the two approaches introduced above we
decided to use Checked C instead of Deputy mainly because the latter is
no longer maintained.

\section{RIOT-POLICE Overview}
\label{sec:riot-police}

RIOT-POLICE is a fork of the RIOT operating system which includes
support for the Checked C language extension. It was initially based on
release \texttt{2018.01} of the RIOT operating system but later updated
to the \texttt{2018.04} release.

The RIOT source code itself is split into various software modules.
Modules which should be included in an application are selected using a
variable in the applications Makefile. The selected modules and the
application code itself are compiled in alphabetic order and linked into
a single freestanding binary. The Checked C toolchain had to be
integrated into this existing build process.

Apart from build system integration, we also had to convert the legacy C
code to Checked C.  Regarding this process our goal was not to convert
the entire RIOT code base since this would (a)~be rather time consuming
and (b)~our assumption was that some parts of the code base (e.g.
drivers) would not profit as much from memory safety as others. Since we
deemed the network stack to be the largest attack vector, we decided to
focus on that.

Within a few months we managed to convert RIOT modules for IPv6
\cite{rfc2460}, UDP \cite{rfc768}, CoAP \cite{rfc7252} and various
utility modules of the network stack to \enquote{checked program scope}
\cite[18\psqq]{checkedc-spec}. Checked C guarantees that no spatial memory
safety violations occur in checked program scope
\cite[4]{tarditi2017checkedc}. Nonetheless, we wanted to continue using
the converted modules from legacy C in order to convert RIOT modules
incrementally. We achieved this through a Checked C language feature
called \enquote{bounds-safe interfaces}. The Checked C compiler inserts
\enquote{implicit conversions between checked types and unchecked types
at bounds-safe interfaces} \cite[91]{checkedc-spec}, thereby ensuring
that calls to checked functions type check even in legacy C code.

Additionally, we wanted to make using Checked C entirely optional. This
was also an objective of the Safe Contiki OS project
\cite[170]{paul2009safecontiki}. Our reasoning behind this was as
follows:

\begin{enumerate}
	\item RIOT supports platforms not supported by LLVM/Clang and
		therefore unsupported by the Checked C compiler.
	\item Some platforms are too constrained for runtime
		overflow checks.
	\item RIOT is a large open source project with many
		contributors. Migrating all developers to a Checked C
		toolchain was not deemed feasible.
\end{enumerate}

We achieved optional safety by defining macros for all new keywords and
types introduced by the Checked C language extension. These macros
fall back to using legacy C types if the \texttt{USE\_CHECKEDC}
preprocessor macro is unset. This allows compilation with a legacy C
complier which doesn't understand the new keywords and types introduced
by the Checked C language extension.

The source code for RIOT-POLICE is available freely on GitHub
\url{https://github.com/beduino-project/RIOT-POLICE}.

\section{Evaluation}

In the following section we will briefly describe the
experience we acquired by integrating Checked C into the RIOT ecosystem.
Furthermore, we are going to evaluate the choices we made regarding this
integration.

\subsection{Optional safety through macros}

The first and most important choice we made is that we wanted to make
using Checked C entirely optional. The reason for this were explained in
\autoref{sec:riot-police}. We considered two different ways of achieving
this:

\begin{enumerate}
	\item Defining macros for new keywords and types, forcing
		developers to use these when writing Checked C code.
	\item Writing a custom pre-processor for Checked C code
		converting it to legacy C code on the fly.
\end{enumerate}

We choose the former because previous research by
\citeauthor{paul2009safecontiki} used the same approach for Safe Contiki OS.
We will evaluate this choice further in the following.  Whether the
costs of achieving optional safety through macros were justified will be
discussed afterwards.  Since \citeauthor{paul2009safecontiki} used
Deputy instead of Checked C they were able to elide Deputy annotations
by defining macros without arguments for Deputy keywords
\cite[170]{paul2009safecontiki}. With Checked C this wasn't possible
since it introduces both new types and keywords. New keywords need to be
stripped and checked types need to be mapped to legacy C types before
compilation with a legacy C compiler is possible. We achieved both by
checking the desired target language in our macro definitions. As an
example consider the definition of the \texttt{ptr} macro in
\autoref{fig:ptr-macro-def}.

\begin{figure}
	\begin{verbatim}
		#ifdef USE_CHECKEDC
		#define ptr(t) _Ptr<t>
		#else
		#define ptr(t) t *
		#endif
	\end{verbatim}

	\caption{Definition of the macro used to declare a pointer to a
		value of a given type.}
	\label{fig:ptr-macro-def}
\end{figure}

The biggest advantage of this macro-based approach is that it doesn't
require any additional changes to the compilation process. Integrating
Checked C into the RIOT build system thus only required changing a few
lines in a single Makefile where the default compiler was set. While
this might seem obvious, it is a huge advantage of Checked C and
shouldn't be taken for granted. We personally made the opposite
experience while attempting to integrate the Rust compiler
into the RIOT buildsystem \cite{tempel2016rust}.

Unfortunately, the macro-based approach also has a significant drawback:
We weren't able to utilize the
\texttt{checked-c-convert} program which is a part of the Checked C
toolchain. This tool allows semi-automating the legacy C to Checked C
conversion process since it \enquote{discovers safely-used pointers
and rewrites them to be checked} \cite[8]{tarditi2017checkedc}. However,
without modifications we wouldn't have been able to use it
even if we had wrote
vanilla Checked C code since it currently doesn't support bounds-safe
interfaces \cite{tempel2018convert}. The reasons why bounds-safe
interfaces were required are explained in \autoref{sec:riot-police}.

It would haven been possible to spend more time on automating the
conversion process by improving this tool. Sadly, we decided against
doing so because we drastically underestimated the amount of work
required
to convert the RIOT modules. The reason why this process was
so laborious is that RIOT makes heavy use of pointers. This is
especially true for protocol parsers like the CoAP implementation. All
of these pointers had to be annotated, since our goal was to convert all
modules to \enquote{checked program scope} \cite[18]{checkedc-spec}.
Converting the network modules themselves wasn't the most laborious
process. The most time was in fact spent writing bound-safe interfaces for
utility modules used by the network modules. This process only required
changes to header files and could have been easily automated since most
of the changes were trivial.

\begin{figure}
	\small
	\begin{verbatim}
		// Bounds-safe interface for fread(3) in Checked C.
		size_t fread(void *p : byte_count(size * nmemb),
		  size_t size, size_t nmemb,
		  FILE *stream : itype(ptr<FILE>));

		// Bounds-safe interface for fread(3) using our macros.
		size_t fread(void *p abyte_count(size * nmemb),
		  size_t size, size_t nmemb,
		  FILE *stream atype(ptr(FILE)));
	\end{verbatim}

	\caption{Declarations of the well-known C library function
		\texttt{fread}: The first using Checked~C and the second using
		our preprocessor macros.}
	\label{fig:macro-fn-prototype}
\end{figure}

Apart from issues related to the conversion process itself, we also
noticed that our macros made the source code less readable. As an
example, consider the two bounds-safe interfaces declared in
\autoref{fig:macro-fn-prototype}. The first bounds-safe interface is
declared using vanilla Checked C, the latter using our macros. In order
to hide Checked C keywords from a legacy C compiler we had to
encapsulate the colon character, used to introduce a Checked C
annotation, in the macro \texttt{abyte\_count}.
In our experience, this makes
it harder to figure out where the annotation starts on first sight.
However, readability issues are not only related to our macros but also
to Checked C itself. For instance, bounds-safe interfaces occasionally
require types to be declared twice. As an example consider the
annotation for the \texttt{stream} argument in
\autoref{fig:macro-fn-prototype}.

It must also be mentioned that our macros might make it harder for new
developers to learn Checked C since they need to learn both vanilla
Checked C and our macros. Adjusting code using our macros without a
basic understanding of Checked C isn't possible. For example, a
developer adjusting the definition in \autoref{fig:macro-fn-prototype}
must be aware that the annotation of the \texttt{stream} argument needs
to be updated when changing its type. Otherwise compilation will fail
but only when compiling with a Checked C compiler, which makes it likely
that these sort of mistakes will go unnoticed.

\subsection{Checked C}

Apart from issues we had with our macro system we also discovered a few
issues with Checked C itself. The biggest issue being related to the
fact that the compiler is \enquote{not recommended for production use}
\cite{checkedccompilerman} yet. Especially the bounds-safe interfaces
feature didn't seem well tested yet. Over the course of a few months we
found eight compiler bugs \cite{david2018incorrecterror,
tempel2018typemismatch, tempel2018checkedfnptr,
tempel2018assumeboundscast, tempel2018castisvalid,
tempel2018nullptrarithmetic, tempel2018incompatibletype,
tempel2018union}.
Four of these were compiler crashes and almost all of
them were related to bounds-safe interfaces. It must, however, also be
noted that all compiler crashes reported by us have been fixed by
the Checked C developers since.

Additionally, it occasionally became obvious that optional safety was
not a Checked C design goal. For example, when converting the CoAP
implementation \texttt{nanocoap} to Checked C we noticed some issues
with the checked pointer type \texttt{nt\_array\_ptr}. This type ensures
that the array referenced by the pointer is always null-terminated. This
property is achieved by initializing it with a null-terminator and
preventing overwrites of the terminator. However, when compiling with a
legacy C compiler the array usually needs to be explicitly initialized
with a null-terminator, for instance using \texttt{memset(\&ary, 0,
sizeof(ary))}. This code won't compile with a Checked C
compiler since it would overwrite the null-terminator.
This is unfortunate since it makes the
conversion process more complicated. It would be desirable for Checked
C to allow overwriting the existing null-terminator with a new one
\cite{david2018ntarryptr}.
As a workaround we used a preprocessor \texttt{\#ifdef} statement to disable these
\texttt{memset(3)} invocations when compiling with a Checked C compiler.

\begin{table}
	\centering
	\begin{tabular}{c r r r}
		Module         & LC (\si{\byte}) & CC (\si{\byte}) & ES (\si{\percent}) \\\hline
		inet\_csum     & 80              & 134             & 68               \\
		netapi         & 266             & 334             & 26               \\
		netreg         & 212             & 370             & 75               \\
		icmpv6\_echo   & 204             & 314             & 54               \\
		icmpv6         & 304             & 516             & 70               \\
		ipv6           & 1462            & 1819            & 24               \\
		pkt            & 16              & 30              & 88               \\
		pktbuf\_static & 1126            & 1530            & 36               \\
		udp            & 588             & 776             & 32              \\\hline
		Total          & 4258            & 5832            & 37
	\end{tabular}

	\caption{Executable size benchmark results.
		\emph{LC/CC:} Module size in bytes when compiled with either
		a legacy C or Checked C compiler.
		\emph{ES:} Increase of the executable size in percent.}
	\label{tbl:executable-size}
\end{table}

Furthermore, the executable size overhead we observed was higher than
expected. We evaluated the overhead in executable size, introduced by the
Checked C language extension, by compiling the RIOT application
\path{examples/nanocoap} for the platform \texttt{pba-d-01-kw2x} with-
and without optional safety features. Afterwards, we compared the size
of the \texttt{.text} sections of the various \texttt{ar(1)} archives
for modules we converted.
The results are shown in \autoref{tbl:executable-size}.

The change in the size of the \texttt{.text} section is difficult to
compare with benchmarks done by the Safe TinyOS and Checked C developers
\cites[2]{cooprider2007tinyos}[10]{tarditi2017checkedc}
due to the fact that different
programs were converted for benchmarking. Additionally, some of the
modules we converted were rather small (e.g. \texttt{pkt}) and made
heavy use of pointers. Nonetheless, the total percentage of observed
overhead in executable size seems rather high with \SI{37}{\percent}
in total, especially compared to the average increase in code size of
\SI{7.4}{\percent} observed by the Checked C Developers
\cite[10]{tarditi2017checkedc} or the \SI{5.2}{\percent} overhead on
average achieved by the Safe TinyOS project
\cite[2]{cooprider2007tinyos}.
The fact that RIOT currently compiles all modules
without link-time optimisation is a probable cause.
Additionally, we didn't optimize our Checked C code in a
way that would allow the compiler to prove more bounds-checks to be
redundant at compile-time.

The fact that the executable size overhead is rather high made it even
more worthwhile to make the safety features entirely optional,
especially due to the fact that RIOT supports quite a few platforms with
tight limits regarding the available code size.

\section{Conclusion}

We successfully integrated Checked C into an existing legacy C ecosystem
and started converting RIOT incrementally from legacy C to Checked C.
Even though not originally intended by the Checked C developers, it
is also possible to retain compatibility with existing legacy C setups
by making Checked C features optional through C preprocessor macros. This makes
it possible to sustain support for IoT devices which are too
constrained for runtime overflow checks. This was especially worthwhile
since the executable size overhead indicated by our benchmarks was
higher than expected.

We consider Checked C a promising technique for improving the security of
existing legacy C software used on constrained devices. Nonetheless, we didn't
propose integrating our changes to RIOT developers because Checked C itself is
still a moving target and the compiler is currently \enquote{not
recommended for production use} \cite{checkedccompilerman}. However,
this can also be seen as an advantage as it allows the Checked C
developers to address some of the issues laid out in this paper.

One of those issues is missing tooling support for bounds-safe
interfaces \cite{tempel2018convert}. Future work should thus focus on
improving tooling for bounds-safe interfaces and our macro system. In
this regard it might be worthwhile to evaluate whether some of the
issues we had with our macro system can be circumvented by writing
vanilla Checked C instead and converting it to legacy C using a custom
C preprocessor. Additionally, future research should focus on further reducing the
executable size overhead we observed in order to also support highly
constrained IoT devices.

	\printbibliography

\end{document}